\documentclass[draft,12pt]{article}
\usepackage{amsmath,amsfonts,palatino,amsthm,epsf}
\newcommand{\beq}{\begin{equation}}
\newcommand{\eeq}{\end{equation}}

\newcommand{\f}{\begin{equation}}
\newcommand{\ff}{\end{equation}}

\begin{document}

\title{Quantum Theory from Quantum Gravity}
\author{
Fotini Markopoulou\thanks{Email address: fotini@perimeterinstitute.ca}\  \
and
Lee Smolin\thanks{Email address:
lsmolin@perimeterinstitute.ca}\\
\\
\\
Perimeter Institute for Theoretical Physics,\\
35 King Street North, Waterloo, Ontario N2J 2W9, Canada, and \\
Department of Physics, University of Waterloo,\\
Waterloo, Ontario N2L 3G1, Canada\\}
\date{November 17, 2003}
\maketitle
\vfill
\begin{abstract}

We provide a mechanism by which, from a background independent model with no quantum mechanics, quantum theory arises in the same limit in which spatial properties appear.  
Starting with an arbitrary abstract graph as the microscopic model of spacetime, our ansatz is that the microscopic dynamics can be chosen so that 1) the model has a low low energy limit which reproduces the non-relativistic classical dynamics of a system of $N$ particles in flat spacetime, 2) there is a minimum length, and 3) some of the
particles are in a thermal bath or otherwise evolve stochastically.
We then construct simple functions of the degrees of freedom of the theory and show that their probability
distributions evolve according to the Schr\"odinger equation.
The non-local hidden variables required to satisfy the conditions of Bell's theorem
are the links in the fundamental graph that connect nodes adjacent in the graph but distant in the approximate
metric of the  low energy limit. In the presence of these
links, distant stochastic fluctuations are transferred into universal quantum fluctuations.

\end{abstract}
\vfill
\newpage



\section{Introduction}

It is often stated that the goal of research in quantum gravity is to find the way in which
nature unifies quantum theory with general relativity.  One way to try to accomplish
this is by a more or less standard quantization of the gravitational field equations,
as in loop quantum gravity\cite{lqg}.  Another is to consider some quantum theory in different
backgrounds, as in string theory\cite{strings}.

In loop quantum gravity, one discovers that the appropriate basis states for
quantum spatial geometry are spin networks, graphs whose edges are labeled by spins.
While a great deal of progress has been achieved, there remain open issues.
One is that the theory is not unique. A second is that to test whether a given
loop quantum gravity theory is
correct, one needs to show how general relativity and flat spacetime arise in the appropriate
limit.  Spin foams, the path-integral evolution of spin networks, arose as a tool
in this effort \cite{spinfoams}.

In many approaches to quantum gravity, the expectation is that the classical space and time
of general relativity are not fundamental but
rather they arise as the low-energy approximate description of the fundamental Planck scale theory.
Quantum theory, on the other hand, is almost invariably expected to hold unmodified all the way down
to Planck distances, although it is often pointed out that a continuous spacetime is already
built in quantum theory.

The question we raise in this paper is whether {\em both} general relativity
and quantum theory may only be approximations to the as-yet unknown
quantum theory of gravity.  We provide a mechanism by which, from a
background-independent model with no quantum mechanics, quantum
theory arises in the same limit in which spatial properties appear.

More specifically, we start with a simple model of the fundamental theory,
based on the adjacency matrix of an abstract graph.
We assume that the underlying model has an approximation
in which classical physics emerges. In particular, we assume that
the theory has a non-relativistic low energy limit in which
some of the nodes of the graph correspond to the positions of particles in three dimensional
space which we call $x^a_i (t)$, where $1=1,...,N$ labels the particles and $a$ is a spatial index
and that the 
$x^a_i (t)$ evolve according to Newtonian dynamics with some potential
$V(|x_i -x_j|)$.  We further ask that {\it some} of the embedding coordinates are subject
to stochastic fluctuations, for example, by being embedded in a heat bath.

We do not discuss here for what choices of dynamics and under what conditions this
classical, low energy limit will emerge. Our goal in this paper is different, it is to show that
quantum mechanics may also appear in the {\em same} limit. In particular,  we show that
quantum theory for all the particles appears only in the limit described above: certain quantities can be defined, functions of both the embedding coordinates and
the original graph, whose probability distributions evolve in time according
to the Schr\"odinger equation.
Planck's constant then turns out to be a derived quantity.

From the perspective of the issues in the foundations of quantum theory,
this is a stochastic hidden variable model.  From the work of Bell \cite{Bell},
we know that any hidden variable theory has to be non-local.  However, in all
these works and the relevant experimental tests, locality is of course defined
with respect to the causal structure of spacetime.  If the smooth 3+1 spacetime
we live in is only approximate, all kinds of possibilities present themselves.
Indeed, in our model, there are two notions of locality. There is a notion of locality
in the graph of the fundamental theory:
two nodes are nearby if they are connected in the graph.  A separate
notion of locality holds in the
embedding of the graph used in the low energy limit.  Two particles,
represented by the embedding
of two nodes in the graph, are nearby if they are close in the metric of the embedding space.

These two notions of locality will not in general coincide. We shall see that the
non-locality needed to derive
quantum theory from a deterministic model arises exactly because of this. The non-locality
required to recover quantum theory from a non-quantum fundamental theory is non-locality in
space. But if space itself is an emergent property, relevant only at a coarse grained level,
the fundamental theory can still be local, if by that we mean local
in the topology in which the fundamental degrees of freedom are defined.
Microscopic locality is a generic property of quantum gravity theories such as spin foams,  in which smooth spacetime is expected to be emergent.

What we present in this article is only an outline of such a mechanism in which quantum theory arises because of the discrepancy between microscopic locality and locality in the emergent spacetime.  
 Much remains to be filled in before one
has a complete model.  We discuss some possibilities and open issues in the Conclusions.

The outline of this paper is as follows.  
In the next section we introduce our model and the precise assumptions about its low energy limit.
The main method of this paper is stochastic differential equations\cite{stochastic},
the elements of
which we need are reviewed in section 3.  We make use of a formulation of quantum
theory due to Nelson, and in section 4 we review the conditions he gives for
a stochastic dynamics to reproduce the predictions of quantum theory\cite{Nelson}.
In section 5 we present the main result of this paper, which
is the derivation of Nelson's conditions from our
model. This is followed by a brief statement of conclusions and questions for
future work.

\section{The model}

We start with the fundamental theory given by a very simple model: a graph $\Gamma$
with a finite set of $N$ nodes. Two nodes can be connected by at most one edge and there are no self-loops (edges from a node to itself).

Two vertices are called {\em adjacent} if they are connected by an edge.
The {\em adjacency matrix} $Q$ of $\Gamma$ is an
$N\times N$ symmetric matrix, with $Q_{ij}=1$ if there is an edge in the graph connecting
nodes $i$ and $j$, and 0 otherwise.  For example, the adjacency matrix of the graph
\[
\begin{array}{c}\mbox{\epsfbox{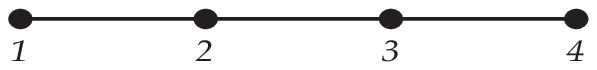}}\end{array}
\]
is 
\[
Q=\left(\begin{array}{cccc}
0&1&0&0\\
1&0&1&0\\
0&1&0&1\\
0&0&1&0
\end{array}\right).
\]

The graph can be thought of as a simple model of a universe of $N$ fundamental building
blocks, where the only information we are given is the adjacency of
these subparts of the universe.

The simplest possible way to set up a correspondence between the graph and a flat 3-dimensional
space is to embed the graph in $R^3$.  The nodes of $\Gamma$ then acquire coordinates
$x_i^a; a=1,2,3; i=1,...,N$.
We will not discuss here the mechanism by which the embedding of the graph into $R^3$ arises.
Such mechanisms have been discussed elsewhere in the spin foam literature\footnote{
For example, in loop quantum gravity and spin foam models, matter fields such as fermion and scalar fields
live on spin network nodes and these are expected to be identified as particles in the low energy limit under
discussion. Alternatively, it may be the case that part of the approximation procedure involves a process
of coarse graining of a spin network or a spin foam, so that what corresponds to
classical spacetime is a coarse grained spin network or spin foam.  If so, again, we
do not need, for the purpose of this paper,to know any details of the coarse graining
procedure except that it takes as input a combinatorial structure and gives, for certain
states or histories required to describe the non-relativistic limit of general relativity, an embedding
of that structure in $R^3$.
Similar remarks apply to other discrete formulations of quantum gravity including causal
sets and dynamical triangulations.   If the theory has a low energy approximation that recovers general relativity, it will
have a further approximation which recovers Newtonian dynamics of point particles in $R^3$.}.

The argument of this paper assumes the existence of that approximation, and asks what becomes of
the information discarded in taking the limit, having to do with the original combinatorial
structure. We shall show that under certain mild assumptions, variables which
are functions of both the embedding variables and the original combinatorial degrees of
freedom evolve quantum mechanically.  We do not need to specify the dynamics of
the model or the process by which Newtonian dynamics is extracted in a low energy and
non-relativistic limit.  We need only a few assumptions concerning the limit,
which we now specify.

We first  require that there is a {\em minimum length} in the embedding, namely that
\f
\left| x^a_i-x^a_j\right|_{\mbox{min}}\sim l,
\ff
$l$ being that minimum distance.  For example, $l$ could be the Planck length.
Let us also call $L$ the {\em average distance} between two nodes:
\f
\left\langle\left(x^a_i-x^a_j\right)^2\right\rangle=:L^2.
\ff

\begin{figure}
\center{
\epsfbox{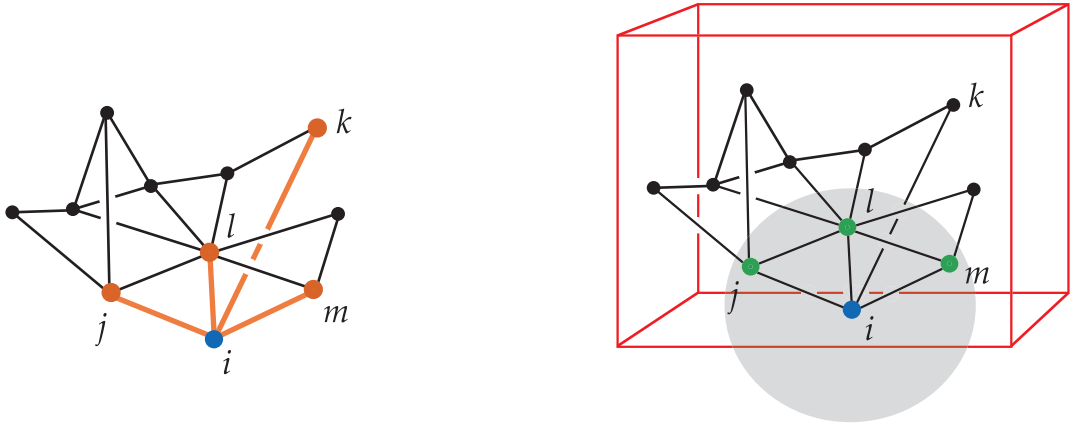}}
\caption{A sketch of how locality in the graph may not be preserved in the embedding.
In the abstract graph on the left, the nearest neighbours of $i$ are $j,k,l$ and $m$.
In the embedded graph on the right, a local neighborhood of $i$ defined by the metric of
$R^3$ is drawn,
and $k$ is far outside it.}
\label{graph2}
\end{figure}

The graph is subject to some microscopic rules of evolution
(that is, the microscopic model is a spin foam with a single in and out graph and presumably
a single interpolating history) such that:

\begin{enumerate}
\item
The model possesses a low energy limit, for $L\gg l$, in
which the node coordinates evolve according to
Newtonian mechanics as if they were massive particles\footnote{It suffices for {\em some} nodes to evolve according to eq.\ (\ref{eq:S}), but for calculational simplicity we take the sum in (\ref{eq:S}) over all $N$ nodes.}:
\f
S_l=\int dt\left(\sum_{i=1}^{N} \frac{m}{2}\left(\dot{x}^a_i\right)^2-V(x)\right).
\label{eq:S}
\ff
For simplicity, we take all the masses to be the same.

\item
{\em Some} of the node positions are subject to a Brownian motion, namely,
in addition to eq.\ (\ref{eq:S}), some of the $\{x_i^a\}$ obey the stochastic
differential equation
\f
d x_i^a(t)=b_i^a(x(t),t) dt+dw_i^a(t),
\ff
where the $dw(t)$ are Gaussian with mean 0, mutually independent, and
\f
\left\langle dw_i^a(t) dw_j^b(t)\right\rangle=
    2{\nu^x}_i dt \delta^{ab}\delta_{ij},
\ff
where $\nu_i$ is the diffusion coefficient for $x_i^a$.   We will give the details
of the fluctuations of the $x$'s in section \ref{x}.

This stochastic part of the evolution of some of the node coordinates could be
due either to the corresponding nodes being in a thermal bath (i.e., there are
hot regions in the embedded model), or some other source of uncertainty in
assigning coordinates to the nodes (i.e., the coordinates could be a coarse-grained
description of several of the underlying nodes).

\item  
There is no requirement that the adjacency of the graph translates into locality
in the embedding.  That is, in the graph the nearest neighbours of node $i$ are the nodes
in $\Gamma$ that can be reached from $i$ by traveling along a single edge.  Similarly, the
next-nearest neighbours are those nodes that can be reached using two edges, and so on.
In the embedding description, given the coordinate $\{x_i^a\}$ of the same node, we may
ask for its neighbours up to distance $r$ from $\{x_i^a\}$, namely all the nodes which
lie inside a ball with center $\{x_i^a\}$ and radius $r$.
Clearly, it is possible for two nodes to be nearest neighbours in the graph
and arbitrarily far apart in the embedding (see fig.\ref{graph2}).

\item
Finally, we will assume for the purposes of this paper that the edges of the graph,
and hence the elements of the adjacency matrix $Q_{ij}$ do not evolve in time.

\end{enumerate}

We may summarize the features of this model as: a fundamental finiteness,
a Newtonian limit in which an external time parameter can be identified
and stochastic fluctuations when the spatio-temporal description is used.
We will see that the very surprising feature is that quantum theory is also
contained in this model, in fact, we will derive it from it!  That is, quantities
can be defined that are simple functions of the graph and the embedding geometry
which, as we will show, satisfy the Schr\"odinger equation.

\section{Stochastic dynamics}

Let us define the matrix
\beq
M^a=X^a+lQ^a
\eeq
where  $Q^a=Q$ and $X^a$ is the $N\times N$ matrix with diagonal elements
$X^a_{ii}=x_i^a$, the positions
of the $N$ particles in the $a$ component, and zeros elsewhere. As before,  $l$ is the minimal
length. The matrix $M$ evolves in time according to
the dynamics in
eq.\ (\ref{eq:S}), which affects only the first term, $X$.

This is the simplest function that contains both the information about
the graph, i.e., the
graph's adjacency matrix, and the information of the embedding of the
graph, i.e., the
coordinates of the nodes.
The  physical intuition is that the eigenvalues
$\lambda_i^a$ of $M^a$ represent corrections to the positions $x_i^a$
arising from the nodes which are adjacent in the graph but non-local with respect to the
embedding.  The motivation for
this stems from the following result, which we will prove in the next
sections:

\begin{quote}
{Let $n$ be the average valence of a node in $Q$.  When the $x^a_i$'s evolve
according to classical mechanics, and when $\sqrt{n} \frac{l^2 }{ L^2} \ll 1$, the
evolution of the
probability distributions for the $\lambda^a_i$'s is given, to leading order in
$\sqrt{n} \frac{l^2 }{L^2}$, by the Schr\"{o}dinger equation.}
\end{quote}

The first step in showing this is to use the standard formulas from perturbation
theory to express $\lambda^a_i$ to leading order as,
\beq
\lambda_i^a=x_i^a+l^2\sum_{j\neq i}\frac{(Q_{ij}^a)^2}{x_i^a-x_j^a}+... .
\label{eq:pert}
\eeq
One can check that the second, fluctuating term in the above expression is of
order
\beq
\Delta \lambda_i^a:=l^2\sum_{j\neq i}\frac{(Q_{ij}^a)^2}{x_i^a-x_j^a}
\sim \sqrt{n}\ \frac{l^2}{L},
\label{eq:lambda}
\eeq
since it is a sum of  $n$ terms of random signs.
Thus,
\beq
\left\langle\frac{\Delta \lambda}{ \lambda}\right\rangle\approx \sqrt{n}\frac{l^2}{L^2 },
\eeq
where $\langle\ \rangle$ means averaged over the nodes and the thermal ensemble.

We will work in the regime
\beq
\sqrt{n} \frac{l^2}{L^2}\ll 1.
\eeq
Note that when $\sqrt{n} \frac{l^2}{L^2} \rightarrow 0$, $M^a \rightarrow X^a$ and
$\lambda^a_i\rightarrow x^a_i$, so that the dynamics becomes classical.
We will study the leading order corrections around this limit.
We will find that quantum mechanics can be understood in this sense as giving the
leading order correction in $\sqrt{n} \frac{l^2}{L^2}$ around the classical limit.

\subsection{The thermal fluctuations of the $\{x^a_i\}$}
\label{x}
Recall that we allow for some of the $\{x_i^a\}$ to be subject to stochastic fluctuations.
The Brownian motion of the $\{x_i^a\}$ is described  by the stochastic differential
equation\cite{stochastic,Nelson}
\f
d x_i^a(t)=b_i^a(x(t),t) dt+dw_i^a(t),
\label{eq:dx}
\ff
for $dt\geq0$,
where the $dw(t)$ are Gaussian with mean 0, mutually independent, and
\f
\left\langle dw_i^a(t) dw_j^b(t)\right\rangle=
    2{\nu^x}_i dt \delta^{ab}\delta_{ij}.
\ff
 $\nu_i$ is the diffusion coefficient for $x_i^a$.  The $dw(t)$ are independent of the
 $x(s)$ for $s\leq t$, so $b_i^a$ is the mean forward velocity
 \f
 b_i^a(x(t),t)=Dx^a_i(t),
 \ff
 where the mean forward derivative $D$ is defined by 
 \f
 Dx(t)=\lim_{\Delta t \rightarrow 0+}\left\langle{\frac{x(t+\Delta t)-x(t)}{\Delta t}}\right\rangle.
 \label{eq:D}
 \ff
 The process is asymmetrical in time, so for $dt\leq 0$ we have
 \f
d x_i^a(t)={b_*}_i^a(x(t),t) dt+d{w_*}_i^a(t),
\ff
where $w_*$ has the same properties as $w$ except that the $dw_*$ are
independent of the $x(s)$ with $s\geq t$.  Similar to the forward case,
\f
{b_*}_i^a(x(t),t)=D_*x_i^a(t)
\ff
is the mean backward velocity, where the mean backward derivative $D_*$ is given by\footnote{
If $x(t)$ is differentiable, then $Dx(t)=D_*x(t)=dx/dt$, but this is not the case
for the thermal motion of the nodes.}
\f
 D_*x(t)=\lim_{\Delta t\rightarrow 0+}\left\langle{\frac{x(t)-x(t-\Delta t)}{\Delta t}}\right\rangle.
 \label{eq:Dstar}
 \ff

We define the {\it average diffusion coefficient} of the $x$'s  to be
\f
\nu^x:=\langle{\nu^x}_i\rangle.
\ff

\subsection{The resulting fluctuations of the $\{\lambda_i^a\}$}

It is not surprising that the thermal fluctuations of the $x$'s lead to fluctuations
of the corrected values, the $\lambda$'s.  It is also not surprising that even if a
particular $x_i$ is not fluctuating, the corresponding $\lambda_i$ does, because of
the second term in eq.(\ref{eq:pert}).  What will be surprising is that the fluctuations
of the $\lambda$'s have a very different character than those of the $x$'s, and this is
the subject of our paper.  Let us first calculate the fluctuations on the corrected positions.

We take the stochastic derivative of the perturbative expansion (\ref{eq:pert}) to find that
\f
d\lambda_i^a=dx_i^a-l^2\sum_{j\neq i}\frac{\left(Q_{ij}^a\right)^2}
            {\left(x_i^a-x_j^a\right)^2} \left(dx_i^a-dx_j^a\right).
\ff
Using (\ref{eq:dx}), we rewrite this as
\f
d\lambda_i^a=\left[b_i^a-l^2\sum_{j\neq i}\frac{\left(Q_{ij}^a\right)^2}
            {\left(x_i^a-x_j^a\right)^2}\left(b_i^a-b_j^a\right)\right]dt
            +dw_i^a
            -l^2\sum_{j\neq i}\frac{\left(Q_{ij}^a\right)^2}
            {\left(x_i^a-x_j^a\right)^2} \left(dw_i^a-dw_j^a\right).
\label{eq:dlambda}
\ff
Thus, the $\lambda$'s evolve by the stochastic differential equation
\f
d\lambda_i^a=\beta_i^a dt+dy_i^a,
\label{eq:l1}
\ff
where $\beta_i^a$ is the mean forward velocity of $\lambda_i^a$,
\f
\beta_i^a(\lambda,t)=D\lambda_i^a,
\label{eq:beta}
\ff
and  the fluctuating part can be read off eq.(\ref{eq:dlambda}) to be
\f
dy_i^a=dw_i^a\left[1-l^2\sum_{j\neq i}\frac{\left(Q_{ij}^a\right)^2}
            {\left(x_i^a-x_j^a\right)^2}\right]+
            l^2\sum_{j\neq i}\frac{\left(Q_{ij}^a\right)^2}
            {\left(x_i^a-x_j^a\right)^2}dw_j^a.
\ff

As in the case of the $x$'s, we may also write
\f
d\lambda_i^a={\beta_*}_i^a dt+{dy_*}_i^a,
\label{eq:betastar}
\ff
where ${\beta_*}_i^a$ is the mean backward velocity of  $\lambda_i^a$
\f
{\beta_*}_i^a(\lambda,t)=D_*{\lambda}_i^a.
\label{eq:l2}
\ff

One reason that some of the $x$'s could be subject to thermal fluctuations is if
in the low energy limit $\sqrt{n} \frac{l^2}{L^2}\ll 1$ there are regions
in the universe at finite temperature.  In this case we may expect that $x$'s that are near
each other in the embedding geometry will be at a similar temperature.
As we shall note in the conclusion, this is not the only possible source of
such local stochastic fluctuations, but it may serve to give intuition.

Let us now relate the diffusion of the $\lambda$'s to that of the $x$'s.
What is of interest in the present paper is nodes whose
$x$'s are not fluctuating.    If a given $x_i$ is not fluctuating, i.e., $\nu_i=0$,
what is the fluctuation of the corrected $\lambda_i$?

The diffusion constant of $\lambda_i$ can be computed from the above equation:
\f
\begin{array}{rl}
  2{\nu^\lambda}_i dt&   =\left\langle\left(d y_i^a\right)^2\right\rangle   \\
  &  =l^4\left\langle\left(\sum_{j\neq i}
        \frac{\left(Q_{ij}^a\right)^2}{\left(x_i^a-x_j^a\right)^2} dw_j^a\right)^2
        \right\rangle\\
  & =l^4\left\langle
    \sum_{j\neq i}
        \frac{\left(Q_{ij}^a\right)^4}{\left(x_i^a-x_j^a\right)^4}2{\nu^x}_j dt
        \right\rangle\\.
\end{array}
\ff
Let us now assume that the graph connects each $\lambda_i^a$ to nodes which are uniformly
distributed in the embedding space. This allows us to deduce that, 
\f
{\nu^\lambda}_i\  = l^4\left\langle
    \sum_{j\neq i}
        \frac{\left(Q_{ij}^a\right)^4}{\left(x_i^a-x_j^a\right)^4}2{\nu^x}_j \right\rangle \\
        \sim n\nu^x\frac{l^4}{L^4}.
        \label{hbar2}
\ff

This tells us that each $\lambda$ is subject to a Brownian motion, {\em even if  its local environment (in the embedding) is at zero temperature.}  This is because the origin of
the brownian motion of the $\lambda_i^a$'s is noise in the $x_i^a$'s which are distant
in space, but near in the graph.  This Brownian motion is {\em irreducible}, in that it cannot be decreased by changing the local conditions, and it is {\em universal} in that it applies to all the $\lambda$'s. We will see in the next section that the consequence of these fluctuations is that the eigenvalues evolve as if quantum theory is true.

\section{Nelson's derivation of quantum mechanics}

We have assumed that  the original
coordinates $x$ are subject to stochastic fluctuations, possibly due to hot regions in the universe. 
We then interpreted the
eigenvalues $\lambda$ of the embedded spin foam, given by the matrix $M$,
 as the corrected positions of the nodes of the graph.  In the previous section we saw that this
results in a Brownian motion for {\em all} of the $\lambda$'s, transferred from the
hot regions to all the $\lambda$'s via edges in the original graph (entries in $Q_{ij}$)
that are non-local connections with respect to the embedding.

Nelson, in his important work \cite{Nelson} considered the stochastic evolution of a
particle in position $x$,  with probability distribution $\rho (x,t)$ and current
velocity $v_i^a(x,t)$, and showed that it will evolve in a way equivalent to a
solution of the time independent Schr\"odinger equation, so long as three conditions
are satisfied.   In this section, we state Nelson's conditions. In the next section,
instead of the position of a particle we consider the corrected positions $\lambda$ of the
nodes of our graph.  We will show that Nelson's equations are satisfied for
the $\lambda$'s in a certain approximation which amounts to specific scaling
relations between $l,L$, the valence of the graph $n$ and the total number of
nodes $N$.  Under these conditions, the nodes evolve according to quantum mechanics!

Nelson's conditions for a particle in position ${\bf x}$, with probability
distribution $\rho ({\bf x},t)$ and current velocity ${\bf v}({\bf x},t)$, are \cite{Nelson}:
\begin{enumerate}
\item
The particle undergoes an irreducible and universal Brownian motion, with
diffusion constant $\nu$ inversely proportional to the mass, $m$.  The
proportionality defines Planck's constant $\hbar$, by
\f
\nu=\frac{\hbar}{m}.
\label{eq:hbar}
\ff
\item
Even though the particle's evolution is fluctuating, its probability density
and current must evolve according to laws that are time reversible.
\item
The current velocity has to be irrotational, namely,  it is proportional to a scalar $S({\bf x})$,
\f
{\bf v}({\bf x},t)=\frac{\partial S({\bf x},t)}{\partial{\bf x}}.
\ff
\end{enumerate}

Nelson shows that, when these conditions are satisfied, one can define the function
\f
\Psi({\bf x},t)=\sqrt{\rho({\bf x},t)}e^{\frac{i}{\hbar}S({\bf x},t)},
\label{Psi}
\ff
and show that the coupled non-linear equations for $\rho$ and ${\bf v}$ reduce to a single linear equation:
\f
i\hbar\frac{d\Psi}{dt}=\left[-\frac{\hbar^2}{2m}\nabla^2+V({\bf x})\right]\Psi.
\label{Sch}
\ff


\section{Satisfying Nelson's conditions}

We will now show that Nelson's conditions are satisfied in our system in the regime 
$\sqrt{n}\frac{l^2}{L^2}\ll 1$. 

\subsection{The kinematics of the $\lambda$'s}

The evolution of the $\lambda$'s in time describes the evolution of the nodes
in the embedded graph.   In place of Nelson's particle at position ${\bf x}$,
we will check for Nelson's conditions for a node $i$ of the graph, its
position given by $\lambda^a_i$.

We have already seen that the $\lambda$'s are subject to an irreducible,
universal Brownian motion, governed by the stochastic
equations (\ref{eq:l1}) and (\ref{eq:betastar}).  We now need to describe the process
in some more detail.  Let us choose a particular
$x^a_i$ whose corresponding $\nu_i=0$.  We can say that that node is ``cold".
We study the probability distribution
$\rho(\lambda,t)$ of the corresponding $\lambda^a_i(t)$.  

First of all, as $\rho$ is a probability,
\f
\int\lambda_i^a\rho(\lambda,t)=1
\ff
must hold.  Next,  as a consequence of (\ref{eq:l1}) and (\ref{eq:betastar}) it follows from the
theory of stochastic differential equations that $\rho$ satisfies
the forward Fokker-Planck equation
\f
\dot\rho=-\frac{\partial\left(\beta_i^a\rho \right)}{\partial\lambda_i^a}
    +\nu^\lambda \nabla \rho,
\ff
and the backward Fokker-Planck equation
\f
\dot\rho=-\frac{\partial\left({\beta_*}_i^a\rho \right)}{\partial\lambda_i^a}
    -\nu^\lambda\nabla\rho.
\ff
The average of the above two equations yields the equation of continuity
\f
\dot\rho=-\frac{\partial \left(\rho v_i^a\right)}{\partial\lambda_i^a},
\label{eq:rhodot}
\ff
where we define $v_i^a$ by
\f
v_i^a(\lambda,t):=\frac{1}{2}\left(\beta_i^a+{\beta_*}_i^a\right).
\ff
From the equation of continuity, we identify $v_i^a$ as the {\em current velocity} \cite{Nelson}.

Subtracting the forward Fokker-Planck from the backward, we find
\f
{\beta_*}_i^a={\beta}_i^a-2\nu^\lambda\frac{1}{\rho}\frac{\partial\rho}{\partial\lambda_i^a}.
\label{eq:fpfp}
\ff
We define the {\em osmotic velocity} $u_i^a(\lambda,t)$ as
\f
u_i^a(\lambda,t):=\frac{1}{2}\left(\beta_i^a-{\beta_*}_i^a\right).
\ff
Then eq.\ (\ref{eq:fpfp}) becomes
\f
u_i^a=\nu^\lambda_i\frac{\partial\ln\rho(\lambda,t)}{\partial\lambda_i^a}.
\label{eq:u}
\ff

\subsection{Conditions 1 and 2}

We now return to the first two of Nelson's conditions.
In postulating (\ref{eq:S}) as the original, uncorrected  evolution in the low-energy limit, we
have assumed that in that limit the nodes behave as if they have the same mass $m$.
Then, to satisfy the first condition above, we simply need to
take eq.(\ref{eq:hbar}) to be the definition of $\hbar$.

We now come to the second condition.  We will follow the way that is implemented by
Nelson in \cite{Nelson}.
The trajectories of the $\lambda$'s are non-differentiable, but one can use the mean
forward and mean
backward derivatives $D$ and $D_*$ defined, in
equations (\ref{eq:D}) and (\ref{eq:Dstar}) to define the average stochastic acceleration
of an eigenvalue $\lambda_i^a$ by
\f
a_i^a(\lambda,t):=\frac{1}{2}\left(DD_*+D_*D\right)\lambda_i^a.
\label{eq:a}
\ff
This definition is time reversible.  Nelson requires that Newton's laws hold for this
averaged acceleration, namely, that it is proportional to the gradient of a potential:
\f
a_i^a=-\frac{1}{m}\frac{\partial V(\lambda)}{\partial\lambda_i^a}.
\label{eq:aV}
\ff

To check this second condition in our model, we will compute the stochastic
acceleration $a$ in eq.(\ref{eq:a}) directly from (\ref{eq:pert}).
The terms we get by computing (\ref{eq:a}) from (\ref{eq:pert}) can be written as
\f
a_i^a=\alpha_i^a+\Delta\alpha_i^a,
\label{eq:aa}
\ff
where 
\f
\alpha_i^a(x,t):=\frac{1}{2}(DD_*+D_*D)x_i^a
\ff
 and $\Delta\alpha_i^a$ is similarly the
symmetrized derivative of the correction term $\Delta \lambda_i^a$ in eq.(\ref{eq:lambda}).

We already have, from eq.(\ref{eq:S}), that
\f
\alpha_i^a(x,t)=-\frac{1}{m}\frac{\partial V(x)}{\partial x_i^a}.
\ff
This can be expanded as
\f
\frac{1}{m}\frac{\partial V(x)}{\partial x_i^a}\sim
    \frac{1}{m}\frac{\partial V(\lambda)}{\partial \lambda_i^a}
    -\Delta\lambda_j^b\frac{1}{m}
    \frac{\partial^2 V(\lambda)}{\partial\lambda_i^a\partial\lambda_j^b}
    +\mathcal{O}\left(\left(\Delta\lambda\right)^2\right).
\label{eq:dlambdaV}
\ff
Recall that $\Delta\lambda_i^a\sim\sqrt{n}l^2/L^2$ (eq.(\ref{eq:lambda})).

For the second term $\Delta\alpha$ in (\ref{eq:aa}), we apply the symmetrized
derivative on (\ref{eq:lambda}).  We use
\f
Df(x,t)= \left ( {d \over dt} + b^{ai} {\partial \over \partial x^{ai}} + \sum_{ai} \nu_i
{\partial^2 \over \partial x^{ai}\partial x_{ai} }
\right ) f(x,t)
\ff
\f
D_* f(x,t)= \left ( {d \over dt} + b^{ai} _* {\partial \over \partial x^{ai}} - \sum_{ai} \nu_i
{\partial^2 \over \partial x^{ai}\partial x_{ai} }
\right ) f(x,t)
\ff
from which it follows that $Dx_i^a=b_i^a$ and $D_*x_i^a=b_{*i}^a$):
\begin{eqnarray}
  \Delta\alpha_i^a&=   &
-\frac{l^2}{2}\sum_{k\neq i}\frac{\left(Q_{ik}^a\right)^2}{\left(x_i^a-x_k^a \right)^2}
        \left(\alpha_i^a-\alpha_k^a\right)
        \nonumber\\
  & &\mbox{}  +2l^2\sum_{k\neq i}
          \frac{\left(Q_{ik}^a\right)^2}{\left(x_i^a-x_k^a \right)^3}
          \left(b_i-b_k\right)^a \left({b_*}_i-{b_*}_k\right)^a
          \nonumber \\
&& + l^2 \sum_{k\neq i} {\nu_k^x  (Q_{ik}^a)^2  \over \left(x_i^a-x_i^a \right)^2 }
\left [
12 {v^a_i - v^a_k \over \left(x_i^a-x_k^a \right)^2} +
2 {\partial_{ak} (v^a_i - v^a_k )\over \left(x_i^a-x_k^a \right)}
+ \partial_{ak}^2 (v^a_i - v^a_k )
\right ]   \
\nonumber \\
&& + 24 \sum_{k\neq i}\frac{ (\nu_k^x )^2  l^2  \left(Q_{ik}^a\right)^2}{\left(x_i^a-x_k^a \right)^5}
\end{eqnarray}

We now make a few physical  assumptions, which allow us to bound the
correction terms. 
First, we assume that the non-local connections are distributed
sufficiently uniformly. This means that for large $n$ there is a single quantity 
 $R$ such, that for all $i$ and $a$,
\f
l^2 \sum_{k \neq i} {Q_{ik}^2 \over (\lambda_i^a - \lambda_k^a )^2} =  Rn{l^2 \over L^2}
[1 + \Delta R_i^a]
\label{second}
\ff
where 
\f
\Delta R_i^a < {1 \over \sqrt{n}}
\label{third}
\ff

As a result, we can estimate, 
\f
 \Delta\alpha_i^a  \sim \sqrt{n}\frac{l^2}{L^2} \left[ \left\langle{\ddot{x}}^a\right\rangle
    +{\left\langle\left(\dot{x}\right)^2\right\rangle \over L}  +
 { \nu^x \sqrt{\left\langle\left(\dot{x}\right)^2\right\rangle} \over L^2} + { (\nu^x)^2 \over L^3}
\right ].
\ff

We now consider the ratio of this to the classical acceleration $\alpha_i^a$.
We have
\f
\frac{\Delta\alpha_i^a }{ \alpha_i^a} \sim \sqrt{n} \frac{l^2}{L^2}
 \left[1 +
{ \left\langle\left(\dot{x}\right)^2\right\rangle \over \ddot{x}^a_i L}
+ { \nu^x \sqrt{\left\langle\left(\dot{x}\right)^2\right\rangle} \over \ddot{x}^a_i L^2} 
+ { (\nu^x)^2 \over \ddot{x}^a_i L^3}
\right].
\label{ratioa}
\ff
However the ratio
\f
{ \left\langle\left(\dot{x}\right)^2\right\rangle \over \ddot{x}^a_i L}
\ff
is on average proportional to twice the ratio of an average kinetic energy to an
average potential energy (because on average the potentials are long ranged forces
so that with $L$ a typical interparticle distance,
$ \ddot{x}^a_i L \sim  V/m$).
Assuming that the system is in equilibrium
we know from the virial theorem of statistical physics that this ratio is
of order unity. 

The remaining terms are order unity or less, assuming that on average
\f
 \nu^x < {    \ddot{x}^a_i L^2 \over  \sqrt{\left\langle\left(\dot{x}\right)^2\right\rangle}  } , \ \ \ \
 (\nu^x)^2  <  \ddot{x}^a_i L^3
\label{moreconditions}
\ff
These tell us that on average the random motion of the $x^a_i$'s is less important
than their bulk motions.  

As a result,  the ratio (\ref{ratioa})  is order $\sqrt{n} \frac{l^2}{L^2}$, so that the
terms in $\Delta \alpha^a_i$ can be neglected compared to $\alpha^a_i$.

As a result we have that to leading order, the stochastic acceleration
is dominated by the classical forces so that,
\f
a^a_i(\lambda , t) \approx - -\frac{1}{m}
\frac{\partial V(\lambda)}{\partial\lambda_i^a}.
\ff
Hence Nelson's second condition is satisfied.

We note that the conditions have been satisfied only to leading order. We expect corrections of order $\sqrt{n} l^2 /L^2$.  We may note that $\hbar$ itself is by eqs. (,\ref{eq:hbar},\ref{hbar2}) proportional to $m \nu^x n l^4 /L^4$.  This tells us that we must assume that $m \nu^x$ is 
large for small $n$, so that $\hbar$ is order unity.

\subsection{Time-independent Schr\"odinger equation}

At this point, and before considering the third condition, we are already able to satisfy the
time-independent Schr\"odinger equation for the simple case in which the probability
distribution is static
\f
\dot\rho=0.
\ff
In this case, condition 3 is trivially satisfied since, by
eq. (\ref{eq:rhodot}),
\f
v_i^a=0.
\ff

We apply $D$ and $D_*$ to $\beta_i^a$ and ${\beta_*}_i^a$ to find that, in the general case,
the stochastic acceleration is given by
\f
a_i^a=\dot u_i^a+v_j^b\frac{\partial v_i^a}{\partial \lambda_j^b}
    -u_j^b\frac{\partial u_i^a}{\partial\lambda_j^b}-\nu_i^\lambda\nabla^2 u_i^a.
\ff
For the static case this reduces to
\f
a_i^a=-u_j^b\frac{\partial u_i^a}{\partial \lambda_j^b}-\nu_i^\lambda\nabla u_i^a.
\ff
In our approximation in which the acceleration is given by eq.\ (\ref{eq:aV}), and
using the definition of $\hbar$ in eq.\ (\ref{eq:hba}),  this can be rewritten as
\f
\frac{\partial}{\partial\lambda_i^a}\left[\frac{1}{2}u^2
        +\frac{\hbar}{m}\frac{\partial u_j^b}{\partial\lambda_j^b}\right]
        =\frac{1}{m}\frac{\partial}{\partial\lambda_i^a}V.
\label{eq:utise}
\ff
We may integrate this to obtain
\f
\frac{1}{2}u^2+\frac{\hbar}{m}\frac{\partial u_j^b}{\partial \lambda_j^b}=\frac{1}{m}\left(V-E\right),
\ff
where $E$ is a constant with dimensions of energy.

Equation (\ref{eq:utise}) is nonlinear in $u$.  However, with a change of 
variables\footnote{Note that with eq.\ (\ref{eq:u})
$\frac{1}{2}u^2+\nu \frac{\partial u_j^b}{\partial \lambda_j^b}= 2\nu^2 \Psi^{-1} \nabla^2 \Psi$.}
\f
\Psi(\lambda)=\sqrt{\rho(\lambda)},
\ff
in eq.\ (\ref{eq:u}), it is equivalent to a linear equation,
the time-independent Schr\"odinger equation
\f
\left[-\frac{\hbar^2}{2m}\nabla^2 +V-E\right]\Psi=0.
\ff
for real $\Psi$.

\subsection{The time dependent Schr\"odinger equation}

We now go on to consider the general case in which the probability
distribution evolves in time. To do this we need to study the third
condition.

We need to show that the current velocity $V^a_i (\lambda )$ has vanishing
curl, to the same order of approximation that the time independent
Schrodinger equation holds
\f
{\partial v^a_i \over \partial \lambda^d_m} -
{\partial v^d_m \over \partial \lambda^a_i} < \sqrt{n} {l^2 \over L^2}.
\label{IIIb}
\ff
We can show that the curl of the current velocity of the $\lambda$'s 
vanishes {\em when} the curl of the {\em classical}  probability current can be
neglected.  The latter quantity is defined as, 
\f
V^a_i (x) = {1 \over 2} \left(b^a_i (x,t) +    {b_*}^{a}_i (x,t)\right).
\ff
We then assume that
\f
{\partial V^a_i \over \partial x^d_m} -
{\partial V^d_m \over \partial x^a_i} < \sqrt{n} {l^2 \over L^2}
\label{eq:neglected}
\ff

Let us start by expressing the current velocity
$v^a_i (\lambda )$ for the eigenvalues in terms of the current
velocity $V^a_i (x)$ for the diagonal elements.  

It is straightforward to show that
\f
v^a_{i}(\lambda )= V^a_{i}(x) -
\sum_{j \neq i} {l^2 Q_{ij}^2 \over (x^a_i-x^a_j)^2} \left(V^a_i (x) - V^a_j (x) \right) + \ldots .
\label{V1}
\ff
Using eq.\ (\ref{eq:pert}), we can write this as
\f
v^a_{i}(\lambda )= V^a_{i}(\lambda ) -
l^2\sum_{j \neq i} { Q_{ij}^2 \over (\lambda^a_i-\lambda^a_j)^2}
\left(V^a_i (\lambda) - V^a_j (\lambda) \right) + \ldots
-l^2\sum_k \sum_{l \neq k} { Q_{kl}^2 \over \lambda^b_k-\lambda^b_l }
{ \partial V^a_i (\lambda ) \over \partial \lambda^b_k}
+ \ldots .
\ff

Under the assumptions (\ref{eq:neglected}-\ref{third},
 it is straightforward to compute the curl of the current velocity for the eigenvalues
\f
{\partial v^a_i \over \partial \lambda^d_m} -{\partial v^d_m \over \partial \lambda^a_i}.
\ff
It contains several terms, but they can all be shown to be smaller than the
terms we neglected in the derivation of the time independent Schr\"{o}dinger equation.
The details of the calculation are given in Appendix A.  
Thus, we have
\f
{\partial v^a_i \over \partial \lambda^d_m} -
{\partial v^d_m \over \partial \lambda^a_i} < \sqrt{n} {l^2 \over L^2}
\label{vtobeshown}
\ff
and we have verified equation (\ref{IIIb}).
Then there exists an $S(\lambda )$ such that
\f
V^a_i = {1 \over m} {\partial S(\lambda ) \over \partial \lambda^a_I }
+ {\cal O} \left ( \sqrt{n} {l^2 \over L^2}  \right )
\label{Sdef}
\ff
By following the same logic as led to the time independent Schr\"{o}dinger equation,
we can follow Nelson's derivation \cite{Nelson}. Using (\ref{Sdef}), we construct the wavefunction
(\ref{Psi}).  Using Nelson's stochastic modification of Newton's laws, which 
follow to order $\sqrt{n} l^2/L^2$, from the conditions we have demonstrated, we can
show that (\ref{Psi}) sastisfies the Schrodinger equation, (\ref{Sch}) up to 
$ {\cal O} \left (\sqrt{n} {l^2 \over L^2} \right )$.

\section{Conclusions}

In this paper,  we started with  a graph-based model of  a quantum theory of gravity which we assumed has a simple kind of low energy limit in which the graph 
is embedded in $R^3$, the graph nodes acquire coordinate positions, and they evolve like Newtonian particles with mass $m$ and potential $V$.  We then showed that in the regime
\f
0<\sqrt{n}\frac{l^2}{L^2} \ll1
\ff
 and when: 1) some of the $x$'s are in thermal equilibrium, so that
 the averaged diffusion coefficient $\nu^x$ is non-vanishing and the conditions of the
 virial theorem are satisfied ($\frac{\langle \dot{x}^2\rangle}{\ddot{x}_i^a }L\sim 1 $), and 2)
the temperature in the region of particle $i$ vanishes, so that
 the diffusion constant due to local influences is zero, then 
the evolution of the
 probability distribution for the eigenvalues $\lambda^a_i$ is
 described by a solution to the time-independent Schr\"odinger equation with
 mass $m$ and potential $V$ and $\hbar$ given by $m\nu_\lambda$.

We have also assumed that the non-local connections are distributed uniformly 
throughout the system. 
This allows us to use eqs's (\ref{second},\ref{third}) as well as to deduce
(\ref{hbar2}).  

Furthermore, under the additional assumption that the curl of the
probability current for the classical variables $V(x^a)$ may be
neglected (eq.\ (\ref{eq:neglected}) ),  the time-dependent Schr\"{o}dinger equation follows, to the same
order of approximation.

We close with a few comments about future work.

The model is not intended to be realistic. Much more work needs to be done
to understand whether or not the basic strategy uncovered here can lead
to a real physical theory.
Among the things that need to be done are:

\begin{enumerate}

\item{}While we show that the Schr\"odinger equation is satisfied to a certain approximation 
by our probability distributions, we do not show that {\it all} solutions to the
Schr\"odinger equation arise this way.   This is due partly to limitations in Nelson's
approach.  For this reason it may be interesting to recast the model in the framework
of Adler's approach to quantum theory as emergent from a dynamics of matrices\cite{Adler}.

\item{}Is quantum mechanics recovered beyond the non-relativisitic
approximation?  Nelson's stochastic quantum theory has been extended to
relativistic field theory \cite{relativeNelson}, so it is possible that
the present results can also be extended.

\item{}The corrections to the approximation in which quantum
theory emerges can  be studied. It is possible that they lead to non-linear
corrections to the  Schr\"odinger equation.  We may note that to keep quantum
effects large compared to terms that have been neglected we must keep
the product $m\nu_x$ large as $\hbar \sim m \nu_x n l^4 /L^4$.

\item{}If we calculate the fluctuations of $\lambda^a_i$'s for whom
the corresponding $x^a_i$'s are subject to thermal noise, is the result
the correct finite-temperature quantum theory?

\item{}The same phenomena may occur in other discrete,
combinatorial theories of quantum gravity, such as causal sets \cite{causal}.
These share with spin foam models the problem that the conditions necessary so
that the information in combinatorial states or histories are well
represented by embedding them in low dimensional manifolds are in general
very hard to satisfy.  The result presented here opens up the possibility
that these conditions do not need to be satisfied. Instead the obstructions
that prevent a good matching between the two notions of locality involved in
embedding a combinatorial structure in a low dimensional manifold so as to
match metric relations may lead instead to the discovery of the source of the
non-local hidden variables necessary for a realistic formulation
of quantum theory.

\item{}Different hypotheses can be considered regarding the origin
of the stochastic fluctions of the $x^a_i$'s. Connecting them with temperature
may, in a more realistic model, lead to predictions of a time dependence of
$\hbar$ that could falsify such a theory. Other possibilities include an intrinsic
uncertainty in the embedding coordinate due to the fact that the embedding is an
emergent property of an underlying fundamental theory whose degrees of freedom
are combinatorial.

\end{enumerate}

\section*{Acknowledgements}

We would like to thank our colleagues at PI and
especially Antony Valentini and Olaf Dreyer for comments and encouragement.
We are grateful also to Steven Adler for detailed comments on the manuscript. 

\appendix

\section{Vanishing of the antisymmetric derivatives of the current velocity}

We show here the details of the calculations that establish (\ref{vtobeshown}).
We begin with the relationship between the two current velocities, $V^a_i$ for
the $x^a_i$ and $v^a_i$ for the $\lambda^a_i$.
\f
v^a_i = V^a_i - l^2 \sum_{k \neq i} 
{Q_{ik}^2 \over (\lambda^a_i -\lambda^a_k )^2 } (V^a_i - V^a_k)
\label{V2}
\ff

We use 
\f
{\partial V^a_i \over \partial \lambda^b_j} = {\partial V^a_i \over \partial x^c_l}
{\partial x^c_l \over \partial \lambda^b_j}
\ff
together with
\f
x^c_l = \lambda^c_l - l^2 \sum_{m \neq l} {Q_{ml}^2 \over \lambda^c_l -\lambda^c_m} + ...
\ff
to find
\f
{\partial V^{a}_i  \over \partial \lambda^b_j} =  {\partial V^{a}_i  \over \partial x^b_j} 
\left ( 1+ l^2 \sum_{m \neq j} {Q_{mj}^2 \over (\lambda^b_j -\lambda^b_m )^2 }
\right )
- l^2 \sum_{l \neq j}   {\partial V^{a}_i  \over \partial x ^b_j} 
 {Q_{jl}^2 \over (\lambda^b_l -\lambda^b_j )^2 }
\ff
Combining this with (\ref{V2}) we have
\begin{eqnarray}
{\partial v^{a}_i  \over \partial \lambda^b_j} & =& {\partial V^{a}_i  \over \partial \lambda^b_j} 
\left [ 1-  l^2 \sum_{k \neq i} {Q_{ik}^2 \over (\lambda^a_li-\lambda^a_k)^2 }
\right ]
l^2 \sum_{k \neq i} {Q_{ik}^2 \over (\lambda^a_i-\lambda^a_k)^2 }
 {\partial V^{a}_k  \over \partial \lambda^b_j} 
\\ \nonumber
&& + 2l^2 \delta^a_b \delta^j_i  
\sum_{k \neq i} 
{Q_{ik}^2 \over (\lambda^a_i -\lambda^a_k )^2 } (V^a_i - V^a_k)
-2 l^2  \delta^a_b 
{Q_{ij}^2 \over (\lambda^a_i -\lambda^a_j )^2 } (V^a_i - V^a_j)
\end{eqnarray}

It is easiest to consider separately the different cases. Beginning with 
$i=j$, we find,
\begin{eqnarray}
{\partial v^{a}_i  \over \partial \lambda^b_i}-{\partial v^{b}_i  \over \partial \lambda^a_i}
&=& {\partial V^{a}_i  \over \partial x^b_i}-{\partial V^{b}_i  \over \partial x^a_i} \nonumber \\
&&-  {\partial V^{a}_i  \over \partial x^b_i } ( l^2 \sum_{k \neq i} Q_{ik}^2  {1 \over ( \lambda^a_i-\lambda^a_k)^2 } ) +
{\partial V^{b}_i  \over \partial x^a_i } ( l^2 \sum_{k \neq i} Q_{ik}^2  {1 \over ( \lambda^b_i-\lambda^b_k)^2 } )   \nonumber \\
&& - l^2 \sum_{k \neq i} Q_{ik}^2  \left [{1 \over ( \lambda^a_i-\lambda^a_k)^2 }
 {\partial V^{a}_k  \over \partial x^b_i }  - 
{1 \over ( \lambda^b_i-\lambda^b_k)^2 }
 {\partial V^{b}_k  \over \partial x^a_i }  
\right ]
\end{eqnarray}
We now impose conditions ({\ref{eq:neglected}-\ref{third}) to find that 
\f
{\partial v^{a}_i  \over \partial \lambda^b_i}-{\partial v^{b}_i  \over \partial \lambda^a_i} 
< \sqrt{n} {l^2 \over L^2} 
\ff
Similarly, for $i\neq j$ and $a \neq b$ we have
\begin{eqnarray}
{\partial v^{a}_i  \over \partial \lambda^b_j}-{\partial v^{b}_j  \over \partial \lambda^a_i}
&=& {\partial V^{a}_i  \over \partial x^b_j}-{\partial V^{b}_j  \over \partial x^a_i} \nonumber \\
&&-  {\partial V^{a}_i  \over \partial x^b_j } ( l^2 \sum_{k \neq i} Q_{ik}^2  {1 \over ( \lambda^a_i-\lambda^a_k)^2 } ) +
{\partial V^{b}_j  \over \partial x^a_i } ( l^2 \sum_{k \neq j} Q_{ik}^2  {1 \over ( \lambda^b_j-\lambda^b_k)^2 } )   \nonumber \\
&& - l^2 \sum_{k \neq i} Q_{ik}^2  {1 \over ( \lambda^a_i-\lambda^a_k)^2 }
 {\partial V^{a}_k  \over \partial x^b_j}  +l^2 \sum_{k \neq j} Q_{jk}^2 
{1 \over ( \lambda^b_j-\lambda^b_k)^2 }
 {\partial V^{b}_k  \over \partial x^a_i }  
\end{eqnarray}
Under the same conditions, it follows directly that,
\f
{\partial v^{a}_i  \over \partial \lambda^b_j}-{\partial v^{b}_j \over \partial \lambda^a_i} 
< \sqrt{n} {l^2 \over L^2} 
\ff
Finally, the reader can check the case $a=b$ and $i\neq j$. After a similar calculation
we find that 
\f
{\partial v^{a}_i  \over \partial \lambda^a_j}-{\partial v^{a}_j \over \partial \lambda^a_i} 
< \sqrt{n} {l^2 \over L^2} 
\ff

\end{document}